\documentclass[letterpaper,aps,prl,twocolumn,showcaps,superscriptaddress,amsmath,amssymb]{revtex4}
\usepackage{graphicx,color,natbib}

\begin{document}
\title{Hunting Mermaids in Real Space: Known Knowns, Known Unknowns and Unknown Unknowns}

\author{C. Patrick Royall}
\affiliation{School of Chemistry, Cantock's Close, University of Bristol, BS8 1TS, UK}
\affiliation{Centre for Nanoscience and Quantum Information, Bristol BS8 1FD, UK}
\affiliation{HH Wills Physics Laboratory, University of Bristol, Bristol BS8 1TL, UK}

\begin{abstract}
We review efforts to realise so-called mermaid (or short-ranged attraction/long ranged repulsion) interactions in 3d real space. The repulsive and attractive contributions to these interactions in charged colloids and colloid-polymer mixtures, may be accurately realised, by comparing particle-resolved studies with colloids to computer simulation. However, when we review work where these interactions have been combined, despite early indications of behaviour consistent with predictions, closer analysis reveals that in the non-aqueous systems used for particle-resolved studies, the idea of summing the attractive and repulsive components leads to wild deviations with experiment. We suggest that the origin lies in the weak ion dissociation in these systems with low dielectric constant solvents. Ultimately this leads even to non-centro-symmetric interactions and a new level of complexity in these systems.
\end{abstract}

\maketitle

\section{Introduction}

Colloids provide important models for liquids and solids, and among their properties that lead to this is the simplicity of their interactions, which may often be treated as being spherically symmetric \cite{ivlev,likos2001}. Perhaps the simplest of these models is the hard sphere, which was famously demonstrated in experiments with sterically stablised colloids \cite{pusey1986}, challenging though it may be realise perfectly hard interactions in practise, as the colloids always carry \emph{some} electrostatic charge \cite{royall2013myth}. Rather earlier than the quasi-hard spheres, the Derjaguin-Landau-Verwey-Overbeek (DLVO) theory \cite{verwey1948} provided the theoretical foundations for a tunable, long-range repulsion between colloids. Soon after, the seminal work of Asakura and Oosawa (AO) showed that, in a solution of non-adsorbing polymers, colloids experience a tuneable attraction due to the polymer degrees of freedom \cite{asakura1954,asakura1958}. These two approaches provide a framework by which attractions, and repulsions, between colloids may be manipulated.

\begin{figure}[tb]
\centering
\includegraphics[width=80 mm]{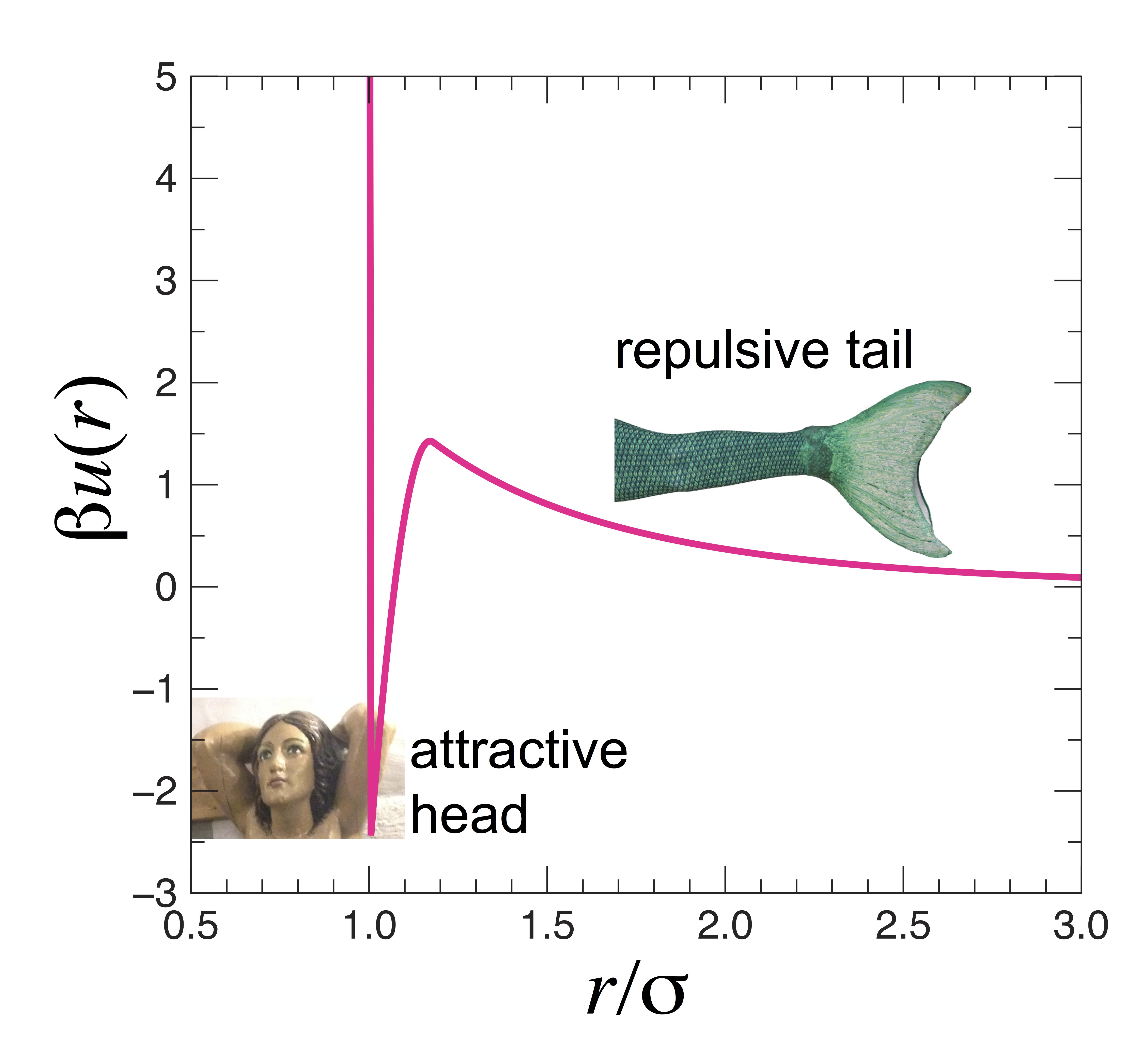} 
\caption{A ``mermaid'', or SALR potential. The ``attractive head'' leads to condensation, but the ``repulsive tail'' opposes this effect, leading to competing interactions and a complex energy landscape \cite{mermaid}.
}
\label{figUMermaid} 
\end{figure}

The combination of both the electrostatic repulsion of DLVO and the AO attractions suggests that colloids may be tuned to have ``mermaid'' interactions, so-called owing to the ``attractive head'' and ``repulsive tail'', Fig. \ref{figUMermaid} \cite{fusco2016}.  Also known as short--ranged attraction--long--ranged repulsion (SALR) systems, these exhibit a rich and exciting phase behaviour, since the competing interactions lead to a complex energy landscape \cite{sciortino2004,tarjus2005,sciortino2005gcs,tarzia2007,ciach2008}. Under such competing interactions, mesophases are predicted, such as lamellae, gyroid phases and clusters \cite{archer2007,ciach2008,ciach2010,zhuang2016,edelman2016}. The clusters formed may themselves order into exotic phases such as cluster crystals \cite{mladek2006,mladek2007,lenz2012} and co-existing cluster fluids \cite{sweatman2014}.

One might imagine then, that given the tuneablity of colloidal systems and that these competing interactions exhibit such a rich phase behavior, then colloidal systems, imaged in the glory of 3d real space \cite{ivlev} would lend themselves to the realisation of the exotic phases thus predicted by simulation and theory. \emph{Yet no ordered phase in a system with competing interactions has ever been observed in 3d real space}, and the reasons underlying this paradox form the subject of this short review. We emphasise the \emph{3d} real space here, because stripe-like lamellar phases and large clusters have indeed been found in 2d systems on an air-water interface, which are well-described by mermaid interactions \cite{ghezzi1997,sear1999,archer2007,law2013} and other approaches, such as using tilted rotating electric fields hold considerable promise \cite{pham2017}. Given that ordered phases have been observed in 2d experiments, here we focus on 3d particle-resolved studies, that is to say work done using confocal microcopy with density- and refractive index-matched systems \cite{ivlev}.

Before proceeding, we note that one of the interesting features of the ``mermaid'' potential is that it may be interpreted as a basic model for ionic systems  \cite{ciach2001}, cement \cite{ioannidou2016} and globular proteins such as lyzozyme \cite{stradner2004,sciortino2004,kowalczyk2010} whose phase behaviour can be compared with colloids \cite{kulkarni2002,stradner2004,cardinaux2007,kowalczyk2010}. We shall therefore make connection to work on protein systems closely connected to the colloids where appropriate.

In no sense is this short review intended to be comprehensive. We have chosen to focus on our own field, particle-resolved studies of colloids. We humbly beg the learned reader for forgiveness if, particularly outside this field, we have neglected to mention relevant work, or indeed if we are to interpret work in a manner that seems at odds with the prevailing view of that field.

This paper is organised as follows. In section \ref{howToMakeAMermaid} we discuss the principles by which well-known interactions between colloids may be tuned such that a mermaid-like potential may be realised. The two contributions -- the attractive head and repulsive tail -- are described in a little more detail in sections \ref{sectionAttractiveHead} and \ref{sectionRepulsiveTail} respectively. In section \ref{sectionParticleResolvedStudies} we outline the experimental systems suitable for realising such an interaction in real space in 3d, using particle-resolved studies. We then review some experiments which set out to measure the two contributions to the  mermaid potential in section \ref{sectionAttractionsAndRepulsionsInParticleResolvedStudies}. Given these contributions, we then consider attempts to realise actual mermaid-type potentials in section \ref{sectionPuttingItAllTogether}. The details of the interactions in some of these papers are considered in section \ref{sectionInteractions}, which leads us to address the observation of \emph{qualitative breakdown} of the idea that one can sum the attractive and repulsive components of a mermaid potential in section  \ref{sectionQualitativeBreakdown}. We present our conclusions and provide some points for future directions in section \ref{sectionConclusions}.

\section{How to make a mermaid: Interactions between colloids}
\label{howToMakeAMermaid}

\subsection{Attractive Head}
\label{sectionAttractiveHead}

As alluded to in the introduction, in order to realise mermaid interactions, one seeks a short-ranged attraction and a long-ranged repulsion. There are many ways to induce interactions between colloids \cite{ivlev,likos2001}. Tuneable attractions which have been implemented in particle-resolved studies of colloids range from depletion/Asakura-Oosawa \cite{dehoog2001}, dipolar (rotating field in 2d) \cite{elsner2009}, critical Casimir \cite{nguyen2013}. Other mechanisms include tuning stablisation against van der Waals attractions \cite{vrij1990}.

In the case of mermaid-type interactions, the Asakura-Oosawa or depletion mechanism has usually been used. 
For polymers that are substantially smaller than the colloids, the resulting mixture can be described by an Asakura-Oosawa (AO) model, which treats the polymer molecules as an ideal gas with hard interactions with the colloids \cite{long1973,vrij1976,dijkstra1999,dijkstra2000}. The AO \emph{effective} interaction potential between two colloids can be written as:
\begin{equation}
\beta u_\mathrm{AO}(r)=\begin{cases}
\text{$\infty$} & \text{for $r<\sigma$}\\
\text{$\frac{\pi(2R_{\rm G})^{3}z_{\rm PR}}{6}\frac{(1+q)^{3}}{q^{3}}$}\times\\
\text{$\{1-\frac{3r}{2(1+q)\sigma}+\frac{r^{3}}{2(1+q)^{3}\sigma^{3}}\}$} & \text{for $\sigma\le r<\sigma$}\\
& +(2R_{\rm G})\\
0 & \text{for $r\ge\sigma+(2R_{\rm G})$}\end{cases}
\label{eqAO}
\end{equation}
where $\beta$ is $1/k_{\rm B}T$. The polymer fugacity $z_{\rm PR}$ is equal to the number density $\rho_{\rm PR}$ of ideal polymers in a reservoir at the same chemical potential as the colloid-polymer mixture. The polymer-colloid size ratio $q=2 R_G/\sigma$ where $R_G$ is the polymer radiius of gyration and $\sigma$ is the colloid diameter.

\subsection{Repulsive Tail}
\label{sectionRepulsiveTail}

Like the mechanisms for attraction noted above, a range of methods have been used in particle-resolved studies of colloids to yield tuneable, long-range repulsions. In addition to electrostatic (DLVO) \cite{verwey1948} repulsions, tuneable magnetic dipolar interactions have been demonstrated \cite{zahn1997}, and electric dipolar interactions are possible (also in 2d). Pertinent to attempts to realise mermaid interactions are the electrostatic interactions. In its linear-Poisson-Boltzmann (DLVO) form, the electrostatic interaction between two colloids takes a Yukawa form.

\begin{equation}
\beta u_\mathrm{yuk}(r)=\begin{cases}
\text{$\infty$} & \text{for $r<\sigma$}\\
\text{$\beta\epsilon_\mathrm{yuk}\frac{\exp(-\kappa(r-\sigma))}{r/\sigma}$} & \text{for $r\ge\sigma$}\end{cases}\label{eqYuk}
\end{equation}
where $r$ is the center--to--center separation of the two colloids. The contact potential is given by 
\begin{equation}
\beta\epsilon_\mathrm{yuk}=\frac{Z^{2}}{(1+\kappa\sigma/2)^{2}}\frac{\lambda_{\rm B}}{\sigma} 
\label{eqbetaepsilon}
\end{equation}

\noindent 
where $Z$ is the colloid charge, $\kappa$ is the inverse Debye screening length and $\lambda_{\rm B}$ is the Bjerrum length. The inverse Debye screening length is given by 
\begin{equation}
\kappa=\sqrt{4\pi\lambda_{B}\rho_\mathrm{ion}}
\label{eqkappa}
\end{equation}

\noindent 
where $\rho_\mathrm{ion}$ is the number density of small monovalent ions. Note that here a (monovalent) salt ion pair would count as two ions.

\section{Particle-resolved studies}
\label{sectionParticleResolvedStudies}

To understand more about how we might realise mermaid-type potentials with particle resolved studies, we need to consider
the particular experimental model systems used. A more detailed discussion may be found in ref. \cite{ivlev} and with a particular focus on the interactions between the particles in ref. \cite{royall2013myth}, so here we briefly note the salient points.

Particle resolved studies uses relatively large colloids (often 3000 nm diameter), so that sedimentation can be a major problem. This means that the particles must be dispersed in a density-matching solvent, which is usually a mixture of two solvents, one with a density larger than and one with a density smaller than the particles. The second requirement is that the solvent has the same refractive index as the colloids, enabling high-resolution 3d optical imaging with confocal microscopy. We note that one elegant means to meet these criteria is to use microgel particles, which are essentially densely cross linked polymers. Like (linear) polymers, these swell, such that the vast majority of the material inside the particle is solvent. This means that good density- and index-matching are intrinsic to the system. However, with the odd notable exception  \cite{alsayed2005}, most work on particle-resolved studies in 3d has focussed on solid particles.

Among systems with solid colloidal particles, those using poly-methyl methacrylate (PMMA) particles (the same material as in the original hard sphere work of Pusey and van Megen \cite{pusey1986}) have dominated the field  \cite{ivlev}. Now the solvents originally used for light-scattering studies which predated particle-resolved studies, and used smaller (typically between 200 and 400 nm diameter) particles had a very low dielectric constant of around two. There was some flexibility of solvent choice, as these particles were small enough that density matching was not required. In any case, it appears that the main deviation of hard-sphere like behaviour came from the steric stabilisation, which induces a slight degree of softness \cite{bryant2002,royall2013myth}.

The larger particles required for particle-resolved studies necessitated density matching solvents, in the form of halogenated solvents such as cyclohexyl bromide. The density-matching solvents typically used have a somewhat larger dielectric constant of for example 5.37 in the case of the density-matching mixture of \emph{cis}-decalin and cyclohexyl bromide  \cite{leunissenThesis,royall2006}. The change in Bjerrum length (the interaction range over which two elementary charges have an energy of interaction equal to the thermal energy $k_BT$) from around 30 nm to 8 nm had significant implications for the degree of ionic dissociation: very little in the older, dielectric constant two solvent based systems, but the newer model systems suitable for particle-resolved studies exhibited enough ion dissociation that the electrostatics, while weak compared to aqueous system were nevertheless strong enough that the phase behaviour exhibited wild deviations from hard spheres, with ``low-density crystals'' at volume fractions $\phi \sim 0.01$ \cite{yethiraj2003,royall2003,leunissenThesis,riosdeanda2015}. In other words, the increase in size of the colloids, to 3000 nm for particle-resolved studies, necessitating the use of a density-matching solvent, led to a fundamental change in the behaviour of the system: the particles exhibited significant repulsions, acting over distances up to tens of microns, ideal for realising mermaid type interactions.

It is worth nothing that even for these density-matching solvents, the dielectric constant can be tuned. While cyclohexyl bromide (and its relative cycloheptyl bromide) remain the most popular, combinations involving carbon tetrachloride \cite{dehoog2001} and tetrachloro ethylene  \cite{ohtsuka2008} lead to density matching solvents with rather lower dielectric constants. The lower dielectric constant would then suppress ion dissociation, leading to a reduction in charging, as exploited by Klix \emph{et al.} (section \ref{sectionQualitativeBreakdown})  \cite{klix2013}.

The nature of the charging in these low-dielectric constant systems is complex and poorly understood \cite{royall2003,leunissenThesis,klix2013}. This leads to behaviour that under certain conditions deviates wildly from expectations. Sadly, we shall see that precisely those conditions required for mermaid-type behaviour, \emph{i.e.} where the particles are close together (short range attraction) and far apart (long-range repulsion) correspond to such deviations from the expected behaviour.

\section{Attractions and Repulsions in Particle-Resolved Studies}
\label{sectionAttractionsAndRepulsionsInParticleResolvedStudies}

\begin{figure}[tb]
\centering
\includegraphics[width=85 mm]{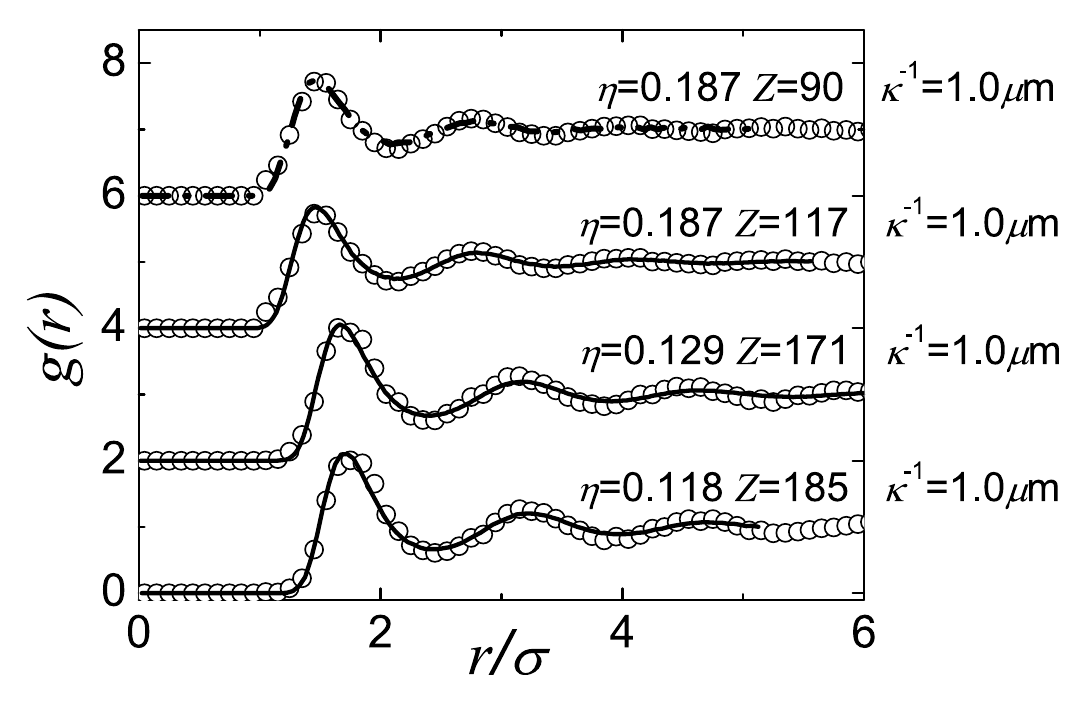} 
\caption{Radial distribution functions for charged colloids at various volume fractions (here denoted $\eta$). Circles are determined from coordinate data from particle-resolved studies. These are compared with simulation data using the Yukawa potential (solid lines) and primitive model (top, dashed). Data offset for clarity \cite{royall2006}.
}
\label{figReentrant} 
\end{figure}

Before we explore successes and, as we shall see, more explicitly, failures, to realise mermaid potentials in real space, let us first consider the components of the interaction -- the attractive head and the repulsive tail. It is possible to measure interactions between colloidal particles and glass walls with total internal reflection microscopy~\cite{bechinger1999,piech2002}, and between pairs of colloidal particles with optical tweezers~\cite{crocker1994}. Optical tweezers were used to measure the AO attraction ~\cite{verma1998}, however the same method \cite{crocker1999} obtained spurious results for the related binary hard sphere system, which also exhibits depletion \cite{roth2000,royall2007jcp}.

While determining the interactions is an important step, demonstrating the potential of a system to exhibit an exotic phase behaviour (presuming it were able to equilibrate) can raise major questions as to whether a system can in fact be described by a simple interaction. In addition to the issues of equilibration (leading often to disordered non-equilibrium states, sometimes termed ``junk'' \cite{poon2002,ramos2005,zaccarelli2007,charbonneau}), the interactions of charged colloids for example are intrinsically density-dependent, due to the fact that the counter-ion contribution to the electrostatic screening term is itself dependent on the colloid concentration, and this effect has been observed in experiments \cite{rojas2002,brunner2002}. More drastic effects can also be observed, likely due to counter-ion condensation leading to unusual phase behaviour in the form of re-entrant melting \cite{royall2006}. Other deviations from the expectations of Eq. \ref{eqYuk} include many-body interactions (\emph{i.e.} a breakdown of pairwise addivity) \cite{brunner2002} and non-centro-symmetric interactions in colloidal crystals \cite{reinke2007}. Other than these last two observations, all of these effects are consistent with the Yukawa interaction, albeit with state-dependent interaction parameters. And significant though the observations of Refs. \cite{brunner2002} and \cite{reinke2007} are, the magnitude of the deviation from Yukawa behaviour is not large.

Attempts to directly compare the results of particle-resolved studies experiments with charged colloids have usually resulted in behaviour consistent with a Yukawa description \cite{brunner2002,royall2003,hsu2005,royall2006}. Such direct comparison typically exploits the fact that, for an isotropic, pairwise additive system, the radial distribution function $g(r)$ is uniquely determined by the pair interaction \cite{hansen}. Such a statement is true in principle, but often in the case of a dense fluid, the $g(r)$ can be rather insensitive to the pair interaction -- an observation than underlies the idea that the hard core (which may be an effective hard core in the case of charged colloids \cite{barker1976}) dominates the structure of such systems \cite{weeks1971}. Nevertheless, under typical experimental conditions (Fig. \ref{figReentrant}), an accuracy of around 20\% is possible in the parameters $\varepsilon_\mathrm{yuk}$ and $\kappa$ that determine the Yukawa interaction Eq. \ref{eqYuk} \cite{royall2003,royall2006}.

Figure \ref{figReentrant} shows the success of the Yukawa description. The parameters obtained are close to those of the primitive model, developed in computer simulation by Vladimir Lobaskin and Per Linse \cite{lobaskin1999} and here implemented by Antti-Pekka Hynninen for a much higher charge asymmetry between ions and colloids \cite{hynninen,hynninen2005}. We see that the parameters in the full primitive model case (top line, Fig, \ref{figReentrant}) is very similar to the value of the Yukawa model (second line down from the top, Fig, \ref{figReentrant}). In short, we conclude that the Yukawa model can provide a good description of the long-ranged repulsion between charged colloids, in particle-resolved studies.

\begin{figure}[tb]
\centering
\includegraphics[width=70 mm]{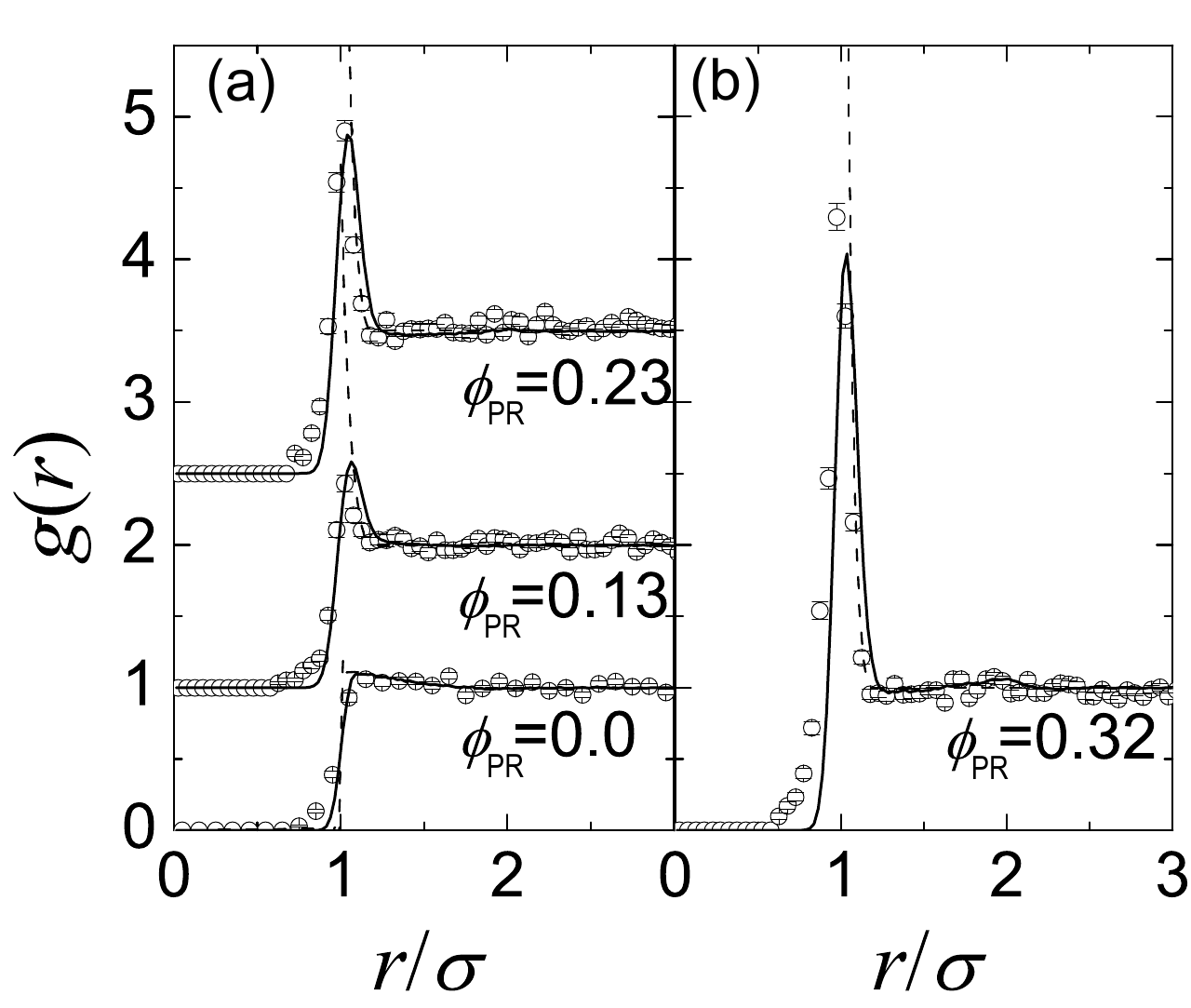} 
\caption{Radial distribution functions $g(r)$ of colloid-polymer mixtures at various polymer concentrations. Monte-Carlo simulations with polymer reservoir volume fraction $\phi_{P}$, according to Eq. (\ref{eqAO}) (solid lines), are compared to the experimental results (circles). Dashed lines correspond to the relation $g(r)\approx\exp(-\beta u_{AO}(r))$. Monte-Carlo simulations consider experimental resolution and polydispersity \cite{royall2007jcp}.}
\label{figGCPArdJCP} 
\end{figure}

In the case of attractions, we can employ the same strategy in the case of a system exhibiting Asakura-Osawa interactions, such as a colloid-polymer mixture \cite{hobbie1998,hobbie1999}. The Asakura-Oosawa interaction is rather shorter-ranged, and many-body effects are small, and indeed vanish if the size ratio $q<0.1547$  \cite{dijkstra1999} and hard to detect for $q=0.25$  \cite{taffs2010}. One issue is that, because the interaction is shorter-ranged, particle tracking errors are more of an issue, so they tend to be comparable in size to the structure of the interaction and thus to the resulting $g(r)$. Such errors can be mimicked by adding Gaussian-distributed noise to coordinates generated by simulation. This leads to a good agreement between experiment and simulation as shown in Fig. \ref{figGCPArdJCP}  \cite{royall2007jcp}. Therefore we conclude that the Asakura-Oosawa model is also well-represented in colloidal systems for particle-resolved studies. It has also been noted that generic short-ranged attractions give similar behaviour \cite{foffi2005}, which has also be seen in the correspondence of the square-well attraction and colloid-polymer mixtures \cite{royall2018,richard2018}.

\section{Putting it All Together: Mermaids in Real Space?}
\label{sectionPuttingItAllTogether}

\begin{figure}[tb]
\centering
\includegraphics[width=60 mm]{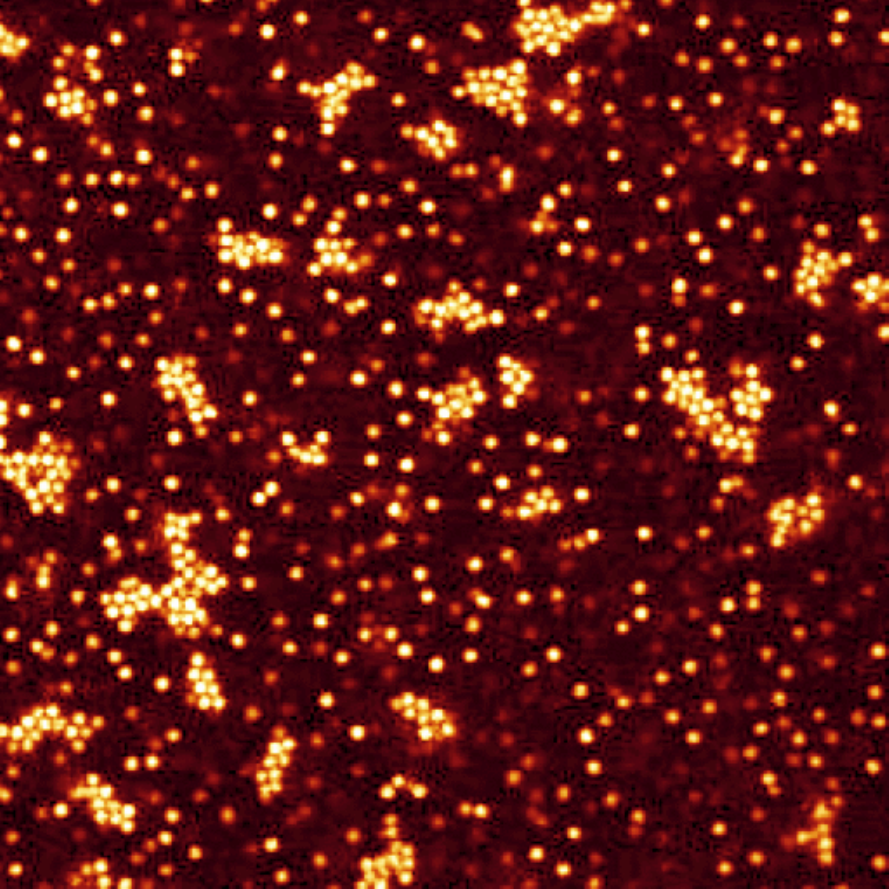} 
\caption{Indications of Mermaid-like behaviour in the form of elongated clusters. 
A confocal micrograph of the clusters in a sample with volume fraction $\phi= 0.086$ and polymer concentration $c_p = 3$ mg cm$^{?3}$ $q=0.021$.
Here the colloids had a diameter of 1320 nm.
Note the spacing between the monomers, indicating a significant strength and range of the repulsive interactions.
Modified with permission from \cite{sedgwick2004}.}
\label{figHelenWilson} 
\end{figure}

We have seen above that it is possible to realise, with reasonable accuracy, the two components of mermaid type interactions - the long-range repulsion and the short-range attraction. Let us now consider what happens when the two are combined -- that is to say, a suspension of charged colloids has polymers added, such that the system exhibits a depletion interaction in addition to the long-ranged electrostatic repulsion.

Early work, particularly that of the Edinburgh group (Fig. \ref{figHelenWilson}) appeared very promising: the colloids were found to cluster, and to form rather elongated clusters, unlike the more spherical clusters that would be expected in the case of systems without the long-ranged repulsions \cite{stradner2004,sedgwick2004}. A little before, inspired by the analogy with atomic nuclei (strong nuclear force versus electrostatic repulsion) such elongated clusters had in fact been \emph{predicted} \cite{groenewold2001,groenewold2004}. Further work followed, with ``Bernal Spirals'' found in a similar system, but one where the colloid concentration was high enough that it percolated, \emph{i.e.} formed a gel \cite{campbell2005}. Simulations using reasonable values for the interactions reproduced the behaviour observed in the experiments, at least qualitatively, in the sense that elongated clusters were formed \cite{sciortino2004,sciortino2005bernal}. It seemed only a matter of time before the particle-resolved studies would deliver ordered phases predicted from simulation, such as lamellae \cite{archer2007}. But no such ordered phase has ever been seen, and we devote the remainder of this article to exploring why this might be.

\section{Interactions in the Mermaid Systems}
\label{sectionInteractions}

We now consider the interaction parameters quoted in experimental realisations of Mermaid-type systems. Campbell \emph{et al.} \cite{campbell2005} report clusters and Bernal spirals in a system in which they measured the colloid charge in a dilute suspension to be $Z = 140$ e per 1.5 $\mu$m diameter colloid, where e is the elementary charge. According to Eq. \ref{eqbetaepsilon}, this maps to a Yukawa contact potential $\beta \varepsilon_\mathrm{yuk} = 35$. Now such a repulsion strength exceeds the kind of attraction strengths typically accessible to the Asakura-Oosawa potential, at least for the polymer-colloid size ratio in question. Using a similar Debye length and reasonable values of the attractive well for the AO attraction, Malins \emph{et al.} found only very limited clustering at $\beta \varepsilon_\mathrm{yuk} = 5$, corresponding to a colloid charge of $Z = 47$ e and expected none at higher Yukawa contact potentials. In another study on gels in systems of competing interactions, Dibble \emph{et al.} \cite{dibble2006} quote a similar value of $Z = 165$ e per $\mu$m diameter colloid. Moreover Sedgwick \emph{et al.} \cite{sedgwick2004} report a charge of $Z < 103$ e in their study of clustering. Although not strictly inconsistent, this seems rather higher than the values for which clustering is expected.

Analysis of these studies paints a picture of anomalously strong repulsions, which would be a struggle for the AO attractions to overcome. One interesting case occurred when the range if the electrostatic interaction was reduced by the presence of salt \cite{royall2005}. While not exactly a mermaid potential as here the repulsions had a comparable range to the attractions, simple addition of Eqs. \ref{eqYuk} and \ref{eqAO} gave an accurate description of the system.  Other than the work of Kohl \emph{et al.} \cite{kohl2016}, which considered similar parameters, as far as we are aware, \emph{no other work has succeeded in finding quantitative agreement with simple addition of the AO attraction and electrostatic repulsion}. We emphasise that this is the \emph{same} system as used in the other experiments (to all intents and purposes, the particular PMMA synthesis run is different, but this does not affect the qualitative behaviour). The only difference is that (presumably) the higher ionic strength corresponding to the shorter Debye length means that there are sufficient ions to suppress the effects we discuss later in section \ref{sectionQualitativeBreakdown}.

\begin{figure}[tb]
\centering
\includegraphics[width=70 mm]{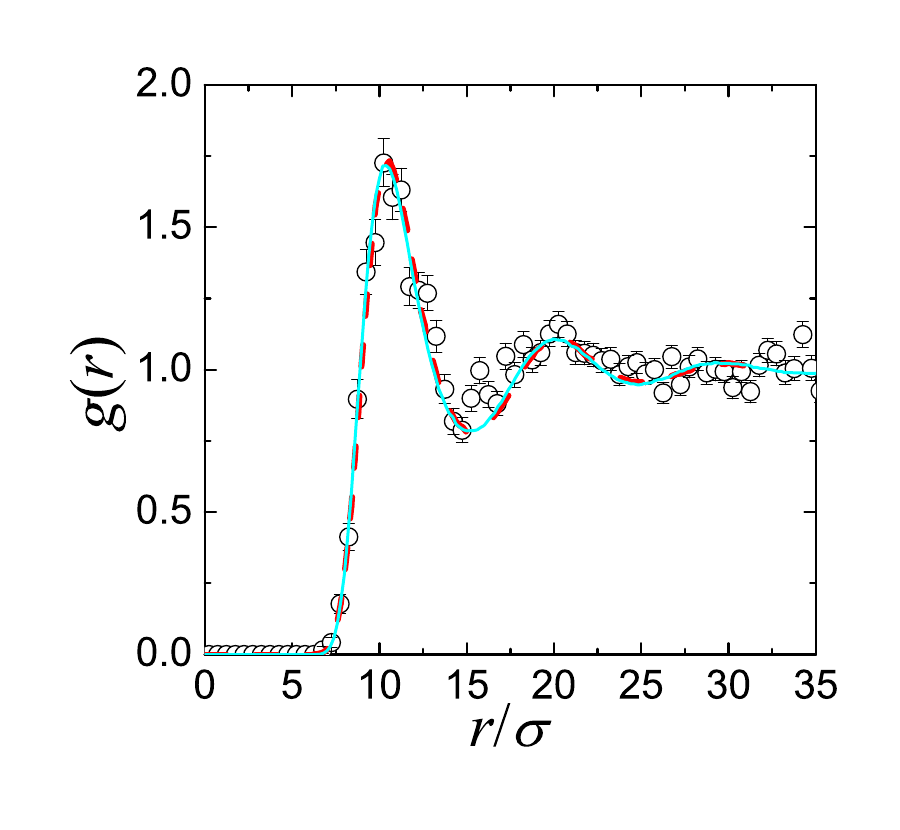} 
\caption{Radial distribution functions. Dashed red line corresponds to a colloid charge $Z=400$ e, solid cyan to $Z=800$ e. We assumed that the Debye screening length was dominated by the colloidal counterions, in other words the system is close to the salt-free limit. This leads to a fitting which depends solely upon $Z$. Lower values of $Z$ gave poor fits, higher values of $Z$ led to crystallization \cite{klix2010}.}
\label{figGFluidChrisPRL} 
\end{figure}

\begin{figure*}[tb]
\centering
\includegraphics[width=170 mm]{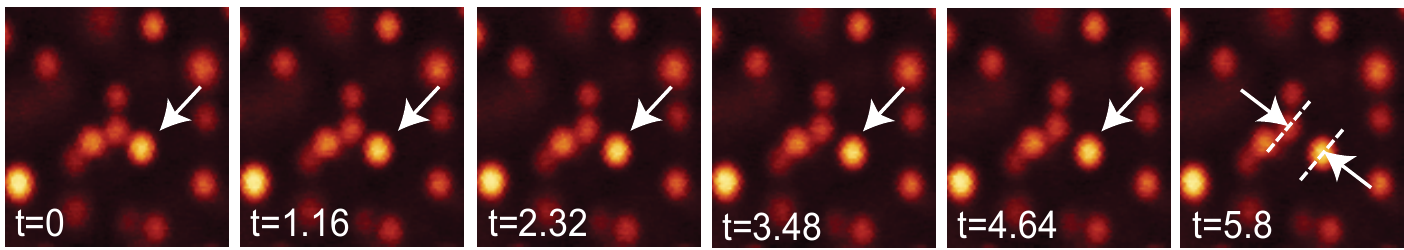} 
\caption{Aging mechanism of a cluster glassy state. An emission process from a 5-membered cluster to a 4-membered cluster, as shown by arrows at a volume fraction of $\phi=0.051$ and polymer concentration $c_p=5.158$ g/l. Time $t$ is expressed in units of the structural relaxation time divided by 1000. Particles are 1.95 $\mu$m in diameter.
\cite{klix2010}.}
\label{figSpittingOutChrisPRL} 
\end{figure*}

Worse was to follow. The numbers quoted above suggest that while the repulsions seem anomalously strong, \emph{i.e.} too strong for clusters to form, the difference was not wild, \emph{i.e.} less than an order of magnitude. This changed with the work of Klix \emph{et al.}  \cite{klix2010}. Figure \ref{figGFluidChrisPRL} shows a $g(r)$ fitted with results from a simulation following the Yukawa model (Eq. \ref{eqYuk}), much like those in Fig. \ref{figReentrant}. However, the volume fraction was very much lower, and requiring neutrality of the overall system (by balancing the colloid charge with counter-ions) places constraints on the Debye length through Eq. \ref{eqkappa}. The estimate for the contact potential was a staggering $\beta \varepsilon_\mathrm{yuk} \approx 1000$. It is hard to imagine how clusters might form in this system, yet, as Fig. \ref{figSpittingOutChrisPRL} shows, indeed polymer-induced depletion interactions nevertheless led to clustering.

Not only did the system cluster (and gel) upon addition of polymer, it aged by emission of particles from the cluster, as Fig. \ref{figSpittingOutChrisPRL} shows. Overall the cluster size throughout the system fell measurably. The electrostatic repulsions held the system in a glassy state, with peculiar sub-diffusive dynamics, even at volume fractions as low as $\phi\approx0.01$ \cite{klix2010}. One possible explanation of this odd behaviour was that somehow the charge was acquired \emph{after} the colloids had clustered or gelled. This would explain the aging behaviour, but it still seems odd that the clusters and gels remained even somewhat intact under such massive electrostatic repulsions.

\section{Qualitative Breakdown of the Yukawa Description: Ion Condensation}
\label{sectionQualitativeBreakdown}

Klix \emph{et al.}  \cite{klix2013} also considered the case when the colloid charge was very weak, comparable to values used in computer simulation studies \cite{sciortino2004,malins2009}. This they effected by tuning the dielectric constant of the solvent to be close to two. Here they again found clustering, and considered each cluster as a separate system, which was shown to be reasonable for the parameters of the system, notably that the interactions between the clusters were small \cite{malins2010}, so the energy landscale of each cluster could be considered in isolation, allowing an analogy to atomic and molecular systems  \cite{wales,doye1995}. With careful mapping of the interaction parameters to computer simulation, Malins \emph{et al.} \cite{malins2009} found that upon increasing the attraction strength, almost all four-membered clusters formed tetrahedra, five-membered triangular biprisms, while six-membered clusters had competing populations of octahedra and polytetrahedra, as also found in experiments on ``sticky spheres'' (with no long-range repulsion) \cite{meng2010}. However in the experiments with the mermaid-type system, the yield of tetrahedra was only 20\%, with the same holding for the triangular biprisms and polytetrahedra \cite{klix2013}.

Conductivity measurements suggested that upon addition of polymer (and clustering), the colloid charge dropped significantly \cite{klix2013}. This observation was consistent with previous work which had shown that the colloid charge drops strongly upon increasing the volume fraction in the absence of polymer, \emph{i.e.} a purely repulsive system \cite{royall2006}, so that one could even imagine the clusters as being locally at high volume fraction (and thus having a lower charge). This is even hinted at in images such as Fig. \ref{figHelenWilson} where the monomers are well-separated, indicating a strong, long-ranged repulsion.

\begin{figure}[tb]
\centering
\includegraphics[width=70 mm]{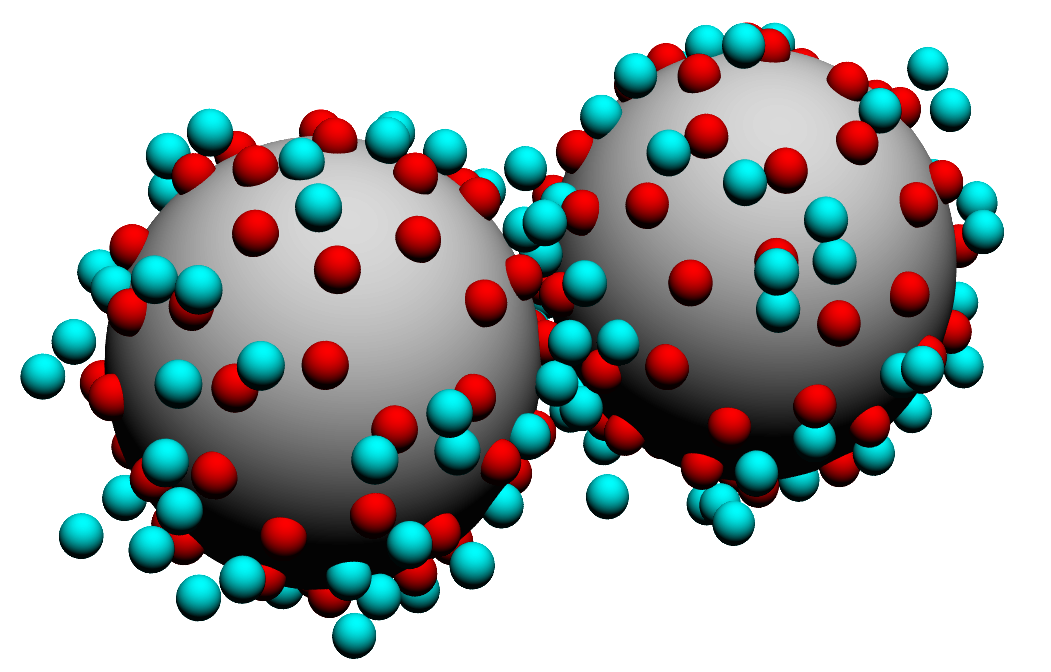} 
\caption{Simulation snapshot of the explicit site Primitive Model. Here the separation between the colloid surfaces is set to $h=0.05\sigma$. Binding sites on the colloid surface and free ions} are shown in red and cyan, respectively (actual size). Note enhanced condensation of ions between colloids.
\cite{klix2013}.
\label{figGoodClusterChrisSciRep} 
\end{figure}

Returning to the work of Klix \emph{et al.}  \cite{klix2013}, noting that the Bjerrum length, at some 23 nm was so large that ions could interact with multiple binding sites on adjacent colloids in a cluster, through an extension of the Primitive model to an explicit site-binding model such that the colloid charge is represented through charging sites on the surface of the particle, the authors suggested that ion condensation between colloids could lead to significant charge asymmetry (Fig. \ref{figGoodClusterChrisSciRep}). This is caused by ions preferentially condensing around contact points between two particles. Such anisotropic ion condensation would lead to a breakdown in the spherically symmetric charge distribution around the colloid implicit in Eq. \ref{eqYuk}. This would then suggest an energy barrier sufficient to prevent the particles forming the tetrahedra (and triangular biprisms and polytetrahedra). This argument required that the colloid dynamics were comparable to those of the small counter- and co-ions. Usually this is absolutely not the case, but the ionic concentration in this system was so low that the time taken for the ions to diffuse their separation was on the 0.1 s timescale, comparable to that of the colloids. Thus the case was made for a breakdown in the Yukawa description in the repulsions preventing the system reaching its ground state, for clusters of four or more particles  \cite{klix2013}.

\section{Discussion and Conclusions}
\label{sectionConclusions}

We have seen that while the components of mermaid potentials, the short range attraction and the long-ranged repulsion can be accurately obtained in 3d particle-resolved studies, their combination remains highly problematic. As described in section \ref{sectionAttractionsAndRepulsionsInParticleResolvedStudies}. The short range attraction is well-captured by colloid-polymer mixtures, while the long-ranged repulsion is found in charged colloids, particularly in the low dielectric constant solvents characteristic of particle-resolved studies, where weak ion dissociation leads to very long Debye screening lengths, up to tens of microns.

However, putting these together leads to a breakdown in the idea that a simple summation of the attractive and repulsive components will describe the system. In most of the work which addresses mermaid-type interactions (section \ref{sectionPuttingItAllTogether}), the colloid-colloid repulsion seems to be anomalously high. In one case the Yukawa contact potential is some 1000 k$_B$T, wildly in excess of that achievable by the Asakura-Oosawa attraction. Furthermore, the large Bjerrum length in some cases can lead to asymmetry in the interactions, \emph{i.e.} a breakdown of the DLVO picture of treating the electrostatic repulsion as a Yukawa interaction. It is our opinion that it is challenging to realise the kind of ordered phases, such as lamellae and gyroid phases, in the systems used for 3d particle-resolved studies, based on polymethyl methacrylate colloids in low dielectric constant solvents.

This observation begs the question of what systems might prove more amenable to such ordering. We noted the early 2d work \cite{sear1999}, in which ordering into lamellae was seen. Now the interactions at interfaces are notoriously complex, and in any case we are mainly interested in 3d systems. One possibility would be use a system where the electrostatic interactions are better understood, for example an aqueous system. However, in aqueous systems, the Debye length is typically much smaller, as the ionic strength is typically much higher. It would be possible to use a solvent of intermediate dielectric constant, in the hope that the Debye length would still be long enough that the electrostatic repulsions could be long-ranged \cite{kodger2015,zhang2016,dong2018}. Alternatively, smaller particles might alleviate the need to match the density, as the sedimentation would be very much reduced by, say, an order of magnitude drop in the particle diameter. This would then perhaps provide a fruitful route to realising the kind of structures predicted for mermaid potentials, controllably, in 3d real space.

\section*{Acknowledgments}

In addition to Per Linse, we dedicate this article also to Antti-Pekka Hynninen, whose brilliance and humour was an inspiration to many. We are grateful to George Stell who coined the term ``mermaid potential''. We thank Patrick Charbonneau for his insightful comments regarding Junk experiments, John Russo for help with the graphics, Bob Evans for his insight into the mermaid state and Paul Bartlett, Alfons van Blaaderen, Alina Ciach, Marjolein Dijkstra, Jeroen van Duijneveldt, Mirjam Leunissen, Wilson Poon, Willem Kegel, Chris Klix, Francesco Sciortino, Richard Sear, Hajime Tanaka and Nigel Wilding for constructive discussions over the years. The author acknowledges the Royal Society, the Japan Society of the Promotion of Science (JSPS), European Research Council (ERC consolidator grant NANOPRS, project number 617266) for financial support and EPSRC grant code EP/H022333/1 for the provision of a confocal microscope, and the Shipwrights Arms for access to a mermaid.


\end{document}